# Realization of a three-dimensional photonic higher-order topological insulator

Ziyao Wang[1†], Yan Meng[2†], Bei Yan[3†], Dong Zhao[1†], Linyun Yang[1], Jing-Ming Chen[1], Min-Qi Cheng[1], Tao Xiao[1], Perry Ping Shum[1], Gui-Geng Liu[4], Yihao Yang[5], Hongsheng Chen[5], Xiang Xi[2*], Zhen-Xiao Zhu[1*], Biye Xie[6*], Zhen Gao[1*]

[1]State Key Laboratory of Optical Fiber and Cable Manufacture Technology, Department of Electronic and Electrical Engineering, Southern University of Science and Technology; Shenzhen 518055, China.

[2]School of Electrical Engineering and Intelligentization, Dongguan University of Technology, Dongguan, 523808, China

[3]Hubei Province Key Laboratory of Systems Science in Metallurgical Process, and College of Science, Wuhan University of Science and Technology, Wuhan 430081, China

[4]Division of Physics and Applied Physics, School of Physical and Mathematical Sciences, Nanyang Technological University, 21 Nanyang Link, Singapore 637371, Singapore.

[5]Interdisciplinary Center for Quantum Information, State Key Laboratory of Modern Optical Instrumentation, ZJU-Hangzhou Global Science and Technology Innovation Center, College of Information Science and Electronic Engineering, ZJU-UIUC Institute, Zhejiang University, Hangzhou 310027, China.

[6]School of Science and Engineering, The Chinese University of Hong Kong, 518172, Shenzhen, China

[†]These authors contributed equally to this work.

*Corresponding author. Email: xix@dgut.edu.cn (X.X); zhuzx@sustech.edu.cn (Z.-X.Z.); xiebiye@cuhk.edu.cn (B.Y.X); gaoz@sustech.edu.cn (Z.G.)

**The discovery of photonic higher-order topological insulators (HOTIs) has significantly expanded our understanding of band topology and provided unprecedented lower-dimensional topological boundary states for robust photonic devices. However, due to the vectorial and leaky nature of electromagnetic waves, it is challenging to discover three-dimensional (3D) topological photonic systems and photonic HOTIs have so far still been limited to two dimensions (2D). Here, we report on the first experimental realization of a 3D Wannier-type photonic HOTI in a tight-binding-like metal-cage photonic crystal, whose band structure matches well with that of a 3D tight-binding model due to the confined Mie resonances. By microwave near-field measurements, we directly observe coexisting topological surface, hinge, and corner states in a single 3D photonic HOTI, as predicted by the tight-binding model and simulation results. Moreover, we demonstrate that all-order topological boundary states are self-guided even in the light cone continuum and can be exposed to air without ancillary cladding, making them well-suited for practical applications. Our work thus opens routes to the multi-dimensional robust manipulation of electromagnetic waves at the outer surfaces of 3D cladding-free photonic bandgap materials and may find novel applications in 3D topological integrated photonics devices.**

Introduction

Recently, the discovery of higher-order band topology [1-3] has revolutionized the study of topological matters with unconventional bulk-boundary correspondence that enables lower-dimensional topological boundary states in at least two dimensions lower than the bulk, in contrast to the conventional first-order topological matters whose topological boundary states only live in just one dimension lower than the bulk. For example, the first-order (higher-order) topological phases host topological edge (corner) states in a two-dimensional (2D) system and topological surface (hinge or corner) states in a three-dimensional (3D) system. To date, extensive studies of various higher-order topological phases have been carried out in condensed matter physics [4-6], photonics [7-31], acoustics [32-57], mechanics [58-60], electric circuits [61-67], and thermal diffusions [68-70].

In general, from the dimensional perspective, higher-order topological phases can be classified into two major classes: 2D [7-23,32-36,58-64,68-70] and 3D [24-31,37-57,65-67]. Compared with the 2D higher-order topological phases whose topological boundary states are limited to first-order one-dimensional (1D) edge states and second-order zero-dimensional (0D) corner states, 3D higher-order topological phases can host first-order 2D surface states, second-order 1D hinge states, and third-order 0D corner states, significantly benefiting multi-dimensional wave manipulation and increasing device integration density. More specifically, there are two different types of 3D HOTIs: 3D HOTIs derived from the generalized Su-Schrieffer-Heeger (SSH) model without quantized multipole moments which are termed Wannier-type 3D HOTIs [37-39], and 3D HOTIs with quantized multipole moments which are dubbed as octupole HOTIs [40-41]. Compared with the octupole HOTIs, the Wannier-type 3D HOTIs do not require negative nearest-neighbor couplings and have been experimentally realized in acoustic crystals [37-39] and electric circuits [65-67] by directly mimicking a 3D SSH model. However, due to the vectorial nature of electromagnetic waves and the lack of mirror reflection symmetry in the vertical direction, the eigenmodes in 3D photonic crystals cannot be simply classified as scalar transverse electric (TE) or transverse magnetic (TM) modes. Consequently, in contrast to the extensive experimental realizations of 3D HOTIs in acoustics [37-57] and electric circuits [65-67], their photonic counterparts have been severely lagged, with only a handful of experimental demonstrations of higher-order Dirac [24] or Weyl [25] semimetals with topological hinge states in 3D photonic crystals. However, due to the absence of complete 3D photonic bandgaps in the photonic higher-order Dirac or Weyl semimetals, the topological hinge states could be scattered into the bulk when encountering obstacles, which inevitably jeopardizes their robust and highly efficient transport. A natural question arises as to whether there exist 3D photonic HOTIs with complete 3D photonic bandgaps and hosting first-order topological surface states, second-order topological hinge states, and third-order topological corner states. Very recently, several 3D photonic HOTIs have been theoretically proposed in tight-binding-like photonic crystals based on the confined Mie resonance [27-30]. However, 3D photonic HOTIs have so far evaded experimental realization in any photonic system.

Here, we report on the first experimental realization of a Wannier-type 3D photonic HOTI in a tight-binding-like photonic crystal, which can be regarded as a photonic realization of the celebrated 3D SSH model. The photonic structure consists of coupled dielectric rods embedded by metallic pillars whose confined Mie resonance serves as artificial atomic orbitals, behaving much like the tight-binding models with nearest-neighbor couplings and exhibiting almost the same band structures. By direct microwave real-space visualization and momentum-space spectroscopy measurements, we experimentally observe coexisting self-guided topological surface states on 2D surfaces, topological hinge states on 1D hinges, and topological corner states confined to 0D corners in a single 3D tight-binding-like photonic crystal without extra claddings, manifesting a full dimensional hierarchy of topological boundary states due to the third-order band topology. Our work experimentally extends photonic HOTIs from 2D to 3D for the first time and may inspire future 3D topological integrated photonic devices and chips.

**Results**

**Design of a Wannier-type 3D photonic HOTIs**

We begin with a 3D SSH model whose unit cell contains eight sites (blue spheres) coupled with nearest-neighbor planar intracell (intercell) couplings $t_w$ ($t_v$) and vertical intracell (intercell) couplings $t_{zw}$ ($t_{zv}$) (pink rods), as shown in Fig. 1a. When the intercell couplings $t_v$ ($t_{zv}$) are larger than the intracell couplings $t_w$ ($t_{zw}$), the 3D SSH model exhibits a Wannier-type third-order topological insulating phase with complete 3D bandgap and coexisting topological surface, hinge, and corner states (see

Supplementary Information) [37-40]. To implement the 3D SSH model in a photonic system, we adopt a 3D tight-binding-like metal-cage photonic crystal (MCPC) [27] whose unit cell is shown in Fig. 1b, where eight dielectric rods (blue rods) are placed between three perforated parallel metallic plates (gold plate) serving as the eight sites and each dielectric rod is surrounded by four metallic pillars (gold rods) to confine the slowly decaying Mie resonances' states, leading to the vectorial electromagnetic waves in 3D photonic crystals being simplified to the scalar-wave-like ones (only TM-like modes) and the 3D MCPC exhibits almost the same scalar-wave-like band structures as those of the 3D tight-binding models. In the context of MCPC, the planar (vertical) intracell and intercell couplings can be modulated by the radii $R_1$ and $R_2$ of the metallic rods and the distance $R$ between the dielectric rods and the central metallic rods (the radius $r_1$ of the air holes and the thickness $h_1$ and $h_2$ of the metallic plates). Since all nearest-neighbor couplings can be modulated independently and flexibly by tuning the geometrical parameters of the MCPCs, a third-order 3D MCPC can be obtained by setting $R_1$ = 1 mm, $R_2$ = 2 mm, $R = 5\sqrt{2}$ mm, $h_1$ = 0.5 mm, $h_2$ = 3 mm, $r$ = 1.6 mm, $r_1$ = 1.5 mm, $d = \frac{3}{10}\sqrt{2}$ mm, $a$ = 15 mm, $a_z$ = 10 mm, respectively. The simulated bulk band structure of the third-order 3D MCPC in the first Brillouin zone (Fig. 1c) is shown in Fig. 1d, which has three complete photonic bandgaps (floral white regions) and bears a striking resemblance to that of the 3D SSH model (see Supplementary Information). To further unveil the third-order band topology of the 3D MCPC, we design a finite 3D MCPC with 4 × 4 × 4 unit cells and calculate its eigenstates spectrum, as shown in Fig. 1e, in which the first-order topological surface states (orange dots) exist in all three photonic bandgaps, the second-order topological hinge states (blue dots) exist in the first and second photonic bandgaps, and the third-order topological corner states (green dots) only emerge in the middle bandgap, exhibiting a multi-dimensional hierarchy of photonic topological boundary states. The simulated $E_z$ field distributions of the topological surface, hinge, and corner eigenstates are shown in Fig. 1f. It can be seen that for the topological surface, hinge, and corner states, their $E_z$ fields are tightly localized at the 2D surfaces, 1D hinges, and 0D corners of the 3D MCPC, respectively.

**Experimental observation of the first-order photonic topological surface states**

Now we start experimentally demonstrating the first-order topological surface states in the 3D MCPC. The fabricated experimental sample is shown in Fig. 2a, which consists of 20 unit cells along the *x* and *z* directions and 4 unit cells in the *y* direction. To show the detailed configuration of each layer, the compositional layers are glided for clarity, as shown in Fig. 2b, where the perforated copper plates with air holes are adopted to induce vertical interlayer and intralayer couplings, and the perforated air foams (white color) are used to fix the metallic and dielectric rods. Note that all six surfaces of the 3D MCPC can support topological surface states and here we only focus on the front (010) surface (parallel to the *x-z* plane). Figure 2d presents the calculated topological surface state dispersions (orange dots) along high-symmetry lines of the projected 2D surface Brillouin zone (Fig. 2c) and the light cone is indicated by the cyan color. It can be seen that the first-order topological surface states exist within all three photonic bandgaps and are completely separated from the bulk bands (grey dots). More interestingly, we find that the topological surface states exhibit a well-defined dispersion even within the light cone continuum, indicating the leakage of topological surface states to the surrounding air is omittable in the whole surface Brillouin zone. This counterintuitive phenomenon fundamentally stems from the strong field localization of the confined Mie resonances [27] and the skyrmion field texture of the topological surface states (see Supplementary Materials) [71], resulting in almost zero overlap with the free-space plane waves and ultralow leakage into the surrounding air. This unique characteristic enables us to observe stable topological surface states in the whole surface BZ, even within the continuum of the light cone. This is nontrivial since conventional photonic topological surface states either need ancillary cladding to prevent the leakage [72-74] or generally have a large radiative loss inside the light cone making them unobservable [75]. We also calculate the quality factors (Q factors) of the topological surface states inside and outside the light cone (see Supplementary

Materials). The high Q factors indicate that the photonic topological surface states of the MCPCs are self-guided without extra confinement by another MCPC or any other cladding.

We then perform experiments to measure the transmission spectra of the topological surface (orange color) and bulk (dark grey) states by placing a source antenna at the center of the front (010) surface and inserting a probe antenna into the surface or bulk, respectively, as shown in Fig. 2e, three photonic bandgaps (floral white color) can be observed as the bulk transmission spectrum exhibits three broad dips in the frequency ranges of 14.2-15.2, 15.3-16.8, and 17-18 GHz, respectively. Within the photonic bandgaps, the surface state transmission spectrum exhibits three sharp peaks, indicating the existence of topological surface states on the 3D MCPC surfaces. To directly observe the field distributions of the topological surface states, we employ a probe antenna to scan the electric field distributions on the 2D surface. The measured field distribution of the topological surface state at 14.6 GHz on the front (010) surface is shown in Fig. 2f, which agrees well with the simulation results shown in Fig. 2g. The field is predominantly localized on the sample's surface, signifying the existence of topological surface states. Upon Fourier-transforming the measured complex field distributions from real space to reciprocal space, we obtain the measured surface state dispersion along the high-symmetry lines $\bar{\Gamma} - \bar{A} - \bar{H} - \bar{K} - \bar{\Gamma}$ in the projected 2D Brillouin zone, as shown in Fig. 2h, the measured results (color map) demonstrate excellent consistency with the simulated surface state dispersion (green circles).

**Experimental observation of the second-order photonic topological hinge states**

Next, we experimentally demonstrate the second-order photonic topological hinge states originating from the higher-order band topology of the 3D MCPC. Note that all twelve hinges support topological hinge states and here we only focus one hinge between (100) and (010) surfaces and simulate its eigenmode dispersion relationship along the $k_z$ direction, as shown in Fig. 3a, in which the surface states, hinge states, bulk states, and light cone are indicated by the orange, blue, grey dots and cyan region, respectively (see Methods for numerical simulation modeling). Similar to the topological surface states, the simulated eigenmode field distribution of the hinge state (Fig. 3b) marked by a red triangle in Fig. 3a shows that the topological hinge state can be well confined at the hinge even if its dispersion is located within the light cone. To excite the topological hinge states, a source antenna (blue stars in Figs. 3d-3e) is placed at the center of the hinge. Subsequently, we use another probe antenna to measure the transmission spectra of the hinge (blue color) and surface (orange color) states, as shown in Fig. 3c, in which the hinge state transmission spectrum (blue color) exhibits two peaks within the two transmission dips of the surface states (orange color). We then map the field distributions on the two surfaces adjacent to the hinge. Fig. 3d shows the measured field distribution of the topological hinge state at 14.9 GHz, matching well with the simulation result shown in Fig. 3e. The field is predominantly localized on the hinge, indicating the presence of topological hinge states. The measured hinge dispersion (color map) along the $k_z$ direction is obtained through Fourier transformation, as shown in Fig. 3f, which shows excellent agreement with the simulation prediction (green circles).

**Experimental observation of the third-order photonic topological corner states**

Finally, we experimentally characterize the third-order photonic topological corner states in the 3D MCPC. Note that all eight corners of the 3D photonic HOTI can support corner states and here we only focus on one single corner. We first measure the transmission spectra of the surface (orange color), hinge (blue color), and corner (green color) states under the excitation of a point source (cyan stars in Figs. 4b-4c) placed near the corner. As shown in Fig. 4a, the corner measurement exhibits a transmission peak within the surface and hinge band gap, which agrees well with the corresponding eigenfrequency ranges of the calculated corner, hinge, and surface eigenstates shown in Fig. 1e. We also plot the measured and simulated field distributions of the corner states at 15.8 GHz in Fig. 4b and Fig. 4c, respectively. The resulting electric field distributions are indeed concentrated at the corner, revealing the tightly localized characteristic of the topological corner states.

**Discussion**

In conclusion, we have experimentally realized a Wannier-type 3D photonic HOTIs with a dimensional hierarchy of coexisting first-order 2D surface states, second-order 1D hinge states, and third-order 0D corner states in a single 3D MCPC. Moreover, we experimentally demonstrate that all-order photonic topological boundary states are self-guided without any ancillary cladding, making them more suitable for practical applications. The ability to integrate 2D surface states, 1D hinge states, and 0D corner states in a single photonic structure may yield potential applications for multi-dimensional photon steering in 3D photonic devices. Since our work provides a versatile platform for implementing a 3D tight-binding model in a 3D photonic crystal, we envision more experimental studies on realizing topological lattice defects such as dislocation [76-78], disclination [79], and Dirac vortex [80-81] in 3D photonic crystals. Along these lines, 3D photonic topological insulators and HOTIs operating at telecom or visible frequencies are on the horizon [82-83]. Besides, 3D artificial gauge fields induced by lattice site deformation [84] and 3D non-Hermitian photonics [31] induced by lattice sites' gain and loss are also readily to be explored based on our structures.

**Method**

**Numerical simulations.** All numerical results presented in this work are simulated by the RF module of COMSOL Multiphysics. The bulk band structure is calculated using a unit cell with periodic boundary conditions in all directions. The perforated copper plates and metallic pillars are modeled as perfect electric conductors (PEC). The surface state dispersion is calculated by adopting a 1 × 4 × 1 supercell and applying periodic boundary conditions along the $x$ and $z$ directions, and open boundary conditions along the $y$ direction. The hinge state dispersion along the $z$ direction is calculated using a 6 × 6 × 1 supercell and applying periodic boundary conditions along the $z$ direction, and open boundary conditions along the $x$ and $y$ directions. In the eigenstate and full-wave simulations of a finite 3D photonic crystal, all six boundaries are set as open boundary conditions.

**Experiment.** The copper plates were fabricated by depositing a 0.035 mm-thick layer of copper onto a substrate of Teflon woven-glass fabric laminate. we adopt perforated dielectric foam (ROHACELL 31 HF with a relative permittivity of 1.04 and a loss tangent of 0.0025) to fix the metallic pillars and dielectric rods. In the experimental measurements, the amplitude and phase of the field are collected by a vector network analyzer (Keysight E5080). The vector network analyzer is connected to two electric dipole antennas, serving as the source and probe, respectively. To excite surface, hinge, and corner states, a source is placed at the surface, hinge, and corner, and a probe is inserted into the air holes to scan the field.

**Data availability**

The data that support the findings of this study are available from the corresponding authors upon reasonable request.

**Code availability**

We use commercial software COMSOL Multiphysics to perform electromagnetic numerical simulations. Requests for computation details can be addressed to the corresponding authors.

**Acknowledgments**

Z.G. acknowledges funding from the National Natural Science Foundation of China under grants No. 62375118, 6231101016, and 12104211, Shenzhen Science and Technology Innovation Commission under grant No. 20220815111105001, and SUSTech under grant No. Y01236148 and No. Y01236248. Y.M. acknowledges the support from the National Natural Science Foundation of China under Grant No. 12304484, Basic and Applied Basic Research Foundation of Guangdong Province under Grant No. 2414050002552, and Shenzhen Science and Technology Innovation Commission under Grant No. 202308073000209. P.S. acknowledges the National Natural Science Foundation of China under grant No. 62220106006, Shenzhen Science and Technology Program under grant No. SGDX20211123114001001.

**Authors Contributions**

Z.G. initiated the project. Z.Y.W., Y.M., B.Y., and X.X. performed the simulations. Z.Y. W., D.Z., Z.X.Z., B.Y.X, and Z.G designed the experiments. Z.G., Z.Y.W., D.Z., X.X., B.Y., L.Y., Y.M., Z.X.Z., and J.M.C. fabricated samples. Z.Y.W., D.Z., and B.Y. carried out the measurements. Z.Y.W., X.X., Z.X.Z., Y.M., B.Y.X, and Z.G. analyzed data. Z.Y.W., X.X. B.Y.X, and Z.G. wrote the manuscript with input from G.G.L, Y.H.Y, P.P.S., and H.S.C. Z.G., X.X, Z.X.Z., and B.Y.X supervised the project.

**Competing Interests**

The authors declare no competing interests.

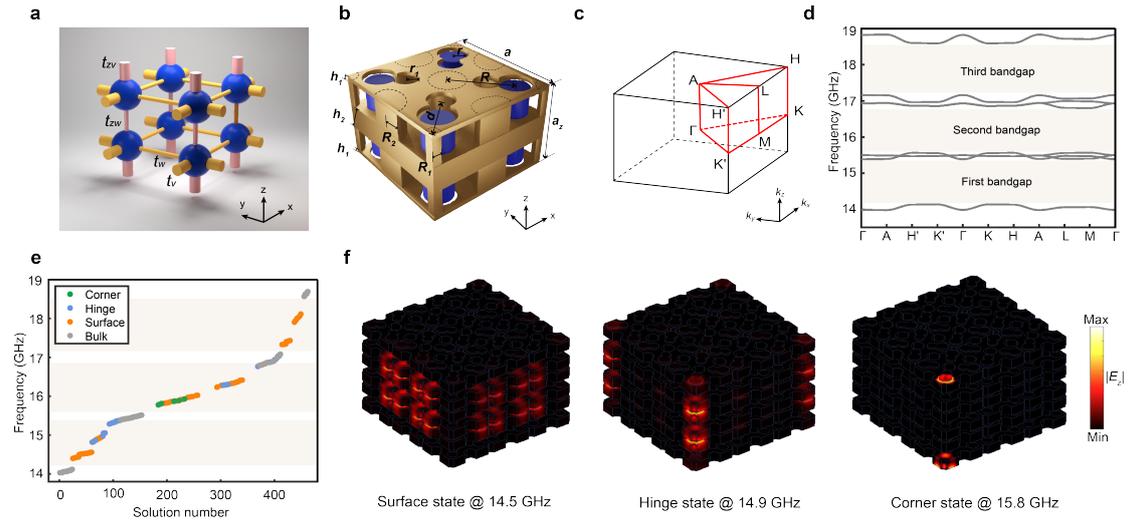

**Fig. 1 | Design of a 3D photonic HOTI. a** Schematic of the 3D SSH model, the blue spheres represent the eight sites, and the pink (yellow) rods represent the vertical (planar) intracell and intercell nearest-neighbor couplings. **b** Unit cell of the 3D MCPC. The golden parts represent the perforated metallic plates and metallic rods, and the blue parts represent the dielectric rods. The lattice constants in the *x-y* plane and *z*-direction are $a$ = 15 mm and $a_z$ = 10 mm, respectively. The other geometrical parameters are $R_1$ = 1 mm, $R_2$ = 2 mm, $R = 5\sqrt{2}$ mm, $h_1$ = 0.5 mm, $h_2$ = 3 mm, $r$ = 1.6 mm, $r_1$ = 1.5 mm, $d = \frac{3}{10}\sqrt{2}$ mm, respectively. **c** 3D Brillouin zone of the MCPC. **d** Simulated bulk band structure of the 3D MCPC along high-symmetry lines, the floral white regions represent the photonic bandgaps. **e** Simulated eigenstate spectrum of a finite 3D MCPC with 4 × 4 × 4 unit cells. **f** Simulated field distributions of the eigenmodes corresponding to the surface, hinge, and corner states, respectively.

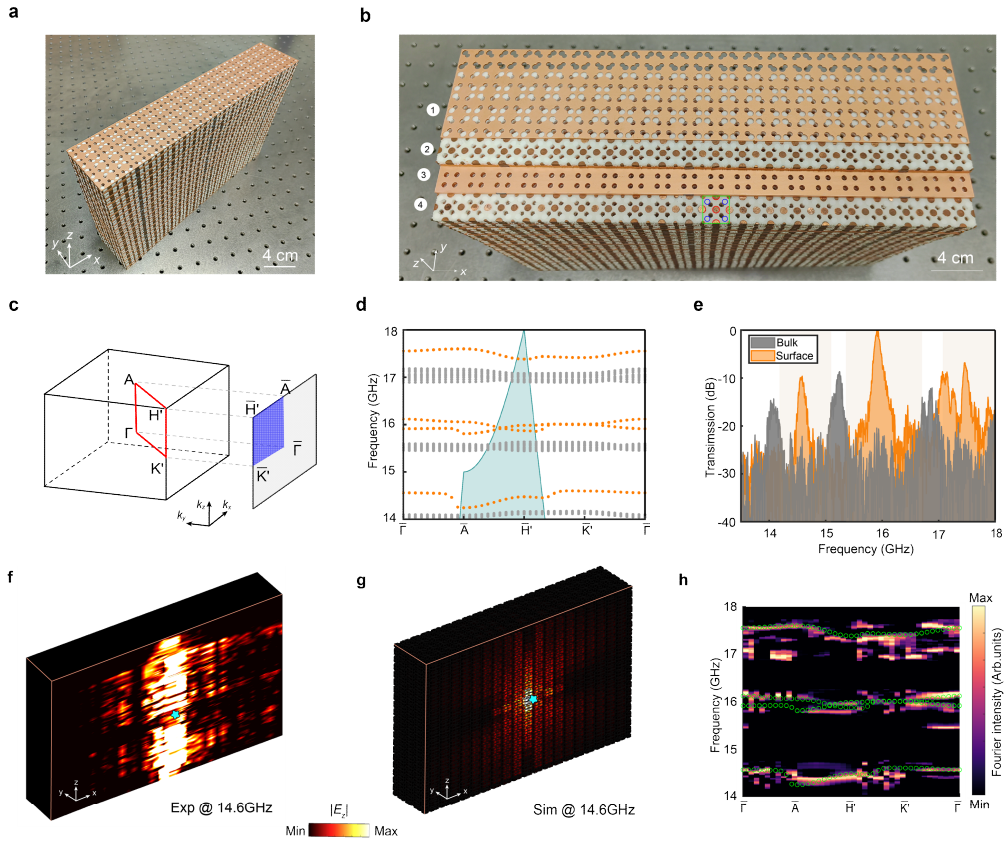

**Fig. 2 | Observation of first-order photonic topological surface states. a** Photograph of the fabricated experimental sample consisting of 4 unit cells in the *y* direction and 20 unit cells in the *x* and *z* directions. **b** The compositional layers are glided for photographing. Each unit cell consists of two perforated copper plate layers and two air foam layers embedded by metallic (red circles) and dielectric (blue circles) rods. **c** Schematic view of the projected 2D surface Brillouin zone. **d** Simulated surface (orange dots) and bulk (grey dots) state dispersions along high-symmetry lines. The cyan region represents the light cone. **e** Measured transmission spectra of the surface (orange color) and bulk (dark grey color) states. The floral white regions represent the photonic bandgaps. **f** Measured electric field distribution of the topological surface states at 14.6 GHz. The cyan star indicates the point source. **g** Simulated electric field distribution of the topological surface states at 14.6 GHz. **h** Measured (color map) and simulated (green circles) surface state dispersions along high-symmetry lines.

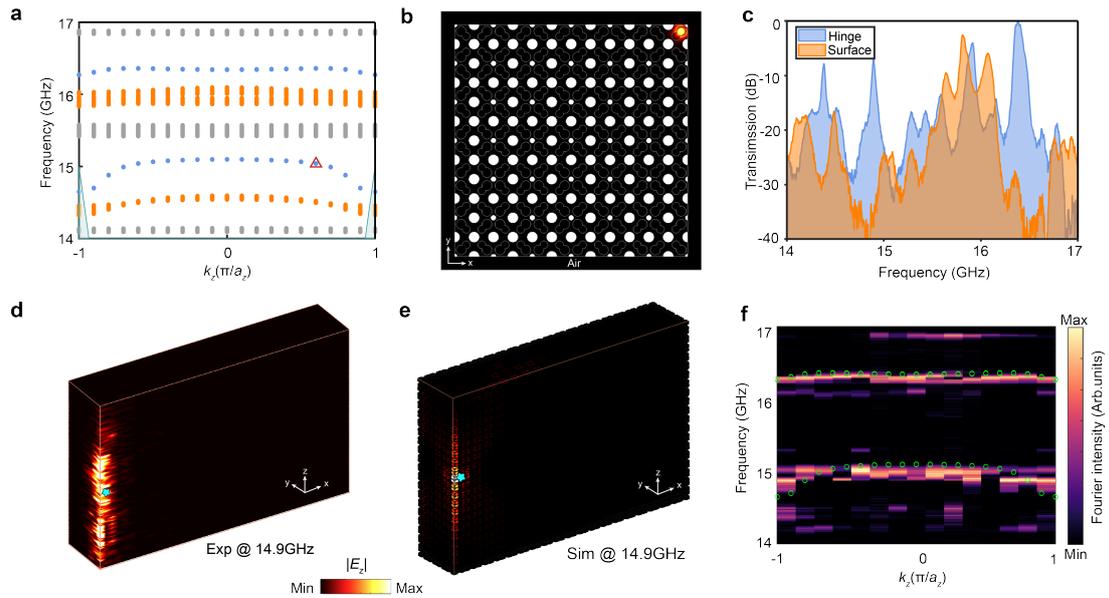

**Fig. 3 | Observation of second-order photonic topological hinge states. a** Simulated bulk (grey dots), surface (orange dots), and hinge (blue dots) state dispersions along the $k_z$ direction. The cyan regions represent the light cone. **b** Simulated electric field distribution of the hinge states corresponding to the red triangle in **a**. **c** Measured transmission spectra of the hinge (blue color) and surface (orange color) states. **d** Measured electric field distribution of the topological hinge states with the source antenna (cyan star) placed at the middle of the hinge. **e** Simulated electric field distribution of the topological hinge states with the point source (cyan star) placed at the middle of the hinge. **f** Measured (color map) and simulated (green circles) hinge state dispersions along the $k_z$ direction.

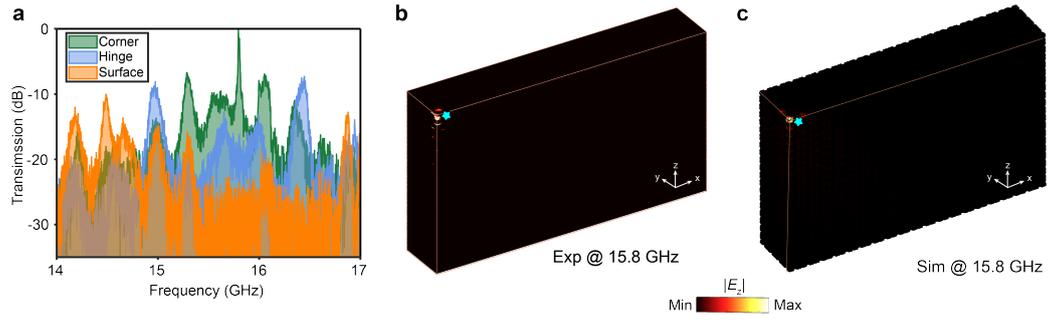

**Fig. 4 | Observation of third-order photonic topological corner states. a** Measured transmission spectra of the surface (orange color), hinge (blue color), and corner (green color) states. **b** Measured electric field distribution of the topological corner states with the source antenna (cyan star) placed at the corner. **c** Simulated electric field distribution of the topological corner states with the point source (cyan star) placed at the corner.